\documentclass[pss]{wiley2sp}
\usepackage{amsmath}
\usepackage{color}

\begin{document}

%
%
%

\title{Electric Transport Properties of the p53 Gene and the Effects of Point Mutations}

\titlerunning{Electric Transport Properties\ldots}

\author{%
  Chi-Tin Shih\textsuperscript{\textsf{\bfseries 1,\Ast}},
  Rudolf A. R\"{o}mer\textsuperscript{\textsf{\bfseries 2}},
  Stephan Roche\textsuperscript{\textsf{\bfseries 3}}}

\authorrunning{Chi-Tin Shih et al.}

\mail{e-mail
  \textsf{ctshih@thu.edu.edu}, Phone
  +886-4-23590121 ext. 3474, Fax +886-4-23594643}

\institute{%
  \textsuperscript{1}\, Department of
Physics, Tunghai University, 40704 Taichung, Taiwan\\
  \textsuperscript{2}\, Department of Physics and Centre for Scientific
Computing, University of Warwick, Gibbet Hill Road, Coventry, CV4
7AL, UK\\
  \textsuperscript{3}\, CEA/DSM/DRFMC/SPSMS/GT, 17 avenue des Martyrs,
38054 Grenoble, France}

\received{XXXX, revised XXXX, accepted XXXX}
\published{XXXX}

\pacs{87.15.Aa, 87.14.Gg, 87.19.Xx}

\abstract{%
%
%
%
\abstcol{%
In this work, charge transport (CT) properties of the p53 gene are
numerically studied by the transfer matrix method, and using
either single or double strand
 effective tight-binding models. A statistical analysis of the
 consequences of known p53 point mutations on CT features is
 performed.}{
 It is found that in contrast to other kind of mutation defects, cancerous mutations result in
 much weaker changes of CT efficiency. Given the envisioned role played by CT in the DNA-repairing mechanism, our theoretical results suggest an underlying physical explanation at the origin of carcinogenesis.}
  }

\maketitle


\section{Introduction}

The electronic transmission properties of DNA molecules are
believed to play a critical role in many physical phenomena taking
place in the living organisms
\cite{EndCS04,Cha07,GutC07,BerKBR04}. For instance, it is believed
that charge transfer (CT) through DNA is inhibited at the damaged sites of the
sequence, owing to misalignements of base pair $\pi$-stacking.
Similarly, base excision repair (BER) enzymes such as Endonuclease
III and MutY are suggested to efficiently locate the DNA base
lesions or mismatches by probing the DNA-mediated CT \cite{RajJB00,YavBSB05,YavSOD06}.

Besides, given that the development of cancers is closely related to the DNA
damage/repair mechanism \cite{LofP96}, the modifications of CT properties when mutations start to develop is therefore an important question to deepen. A most important gene in  cancer research is $p53$ also known as the ``guardian of the genome'' \cite{Lan92}. Indeed, p53
encodes the tumor suppressor $TP53$ protein that suppresses the
tumor development by activating the DNA repair mechanisms or the
cell apoptosis process if DNA reparation is impossible. There
are 20303 base pairs in the $p53$ sequence (NCBI access number
X54156). More than $50\%$ of human cancers are related to the
mutations of the $p53$ gene which usually jeopardize the efficient
activity of $TP53$ \cite{She04}. Most of the cancerous
mutations are point mutations
--- a base pair substituted by another --- with distributions
along the DNA sequence that are highly non-uniform
\cite{PetMKI07}. Each point mutation can be described by two
parameters $(k,s)$, respectively giving the mutation position $k$
on the sequence and the nucleotide type $s$ (either A, C, G, or T) substituting the
original one. The most frequent mutation locations
found in the cancer cells are named mutation ``hotspots''.
 From the International Agency for
Research on Cancer (IARC) database \cite{PetMKI07}, it is found
that most hotspots of $p53$ are located in the exons $5\sim 8$ in
the interval from the $13055$th to the $14588$th nucleotide. The
$13203$th base pair has the highest frequency of occurrence (1055
times) and more than $80\%$ of the total 23544 cases in the
database occur on $1\%$ of the base pairs of the $p53$. The
mutation $(k,s)$ is said to be ``cancerous'' (``noncancerous'') if
it is (not) found in the IARC database.

In this paper, the effects of all possible point mutations on CT
are studied for the p53 gene using appropriate tight-binding
models and energy parameters which are know to reproduce experimental
results or first principle calculations \cite{CunCPD02,Sta05}. We
find that anomalously small changes of CT efficiency
modulations coincide with cancerous mutations. In contrast,
non-cancerous mutations result, on average, in much larger changes
of the CT properties. From this analysis, we propose a new
scenario for understanding the underlying origin of how cancerous
mutations shortcut the DNA damage/repair processes.


\section{Models for charge transport in DNA}
\label{sec:model}

\begin{figure}
  (a)\includegraphics[width=.21\textwidth]{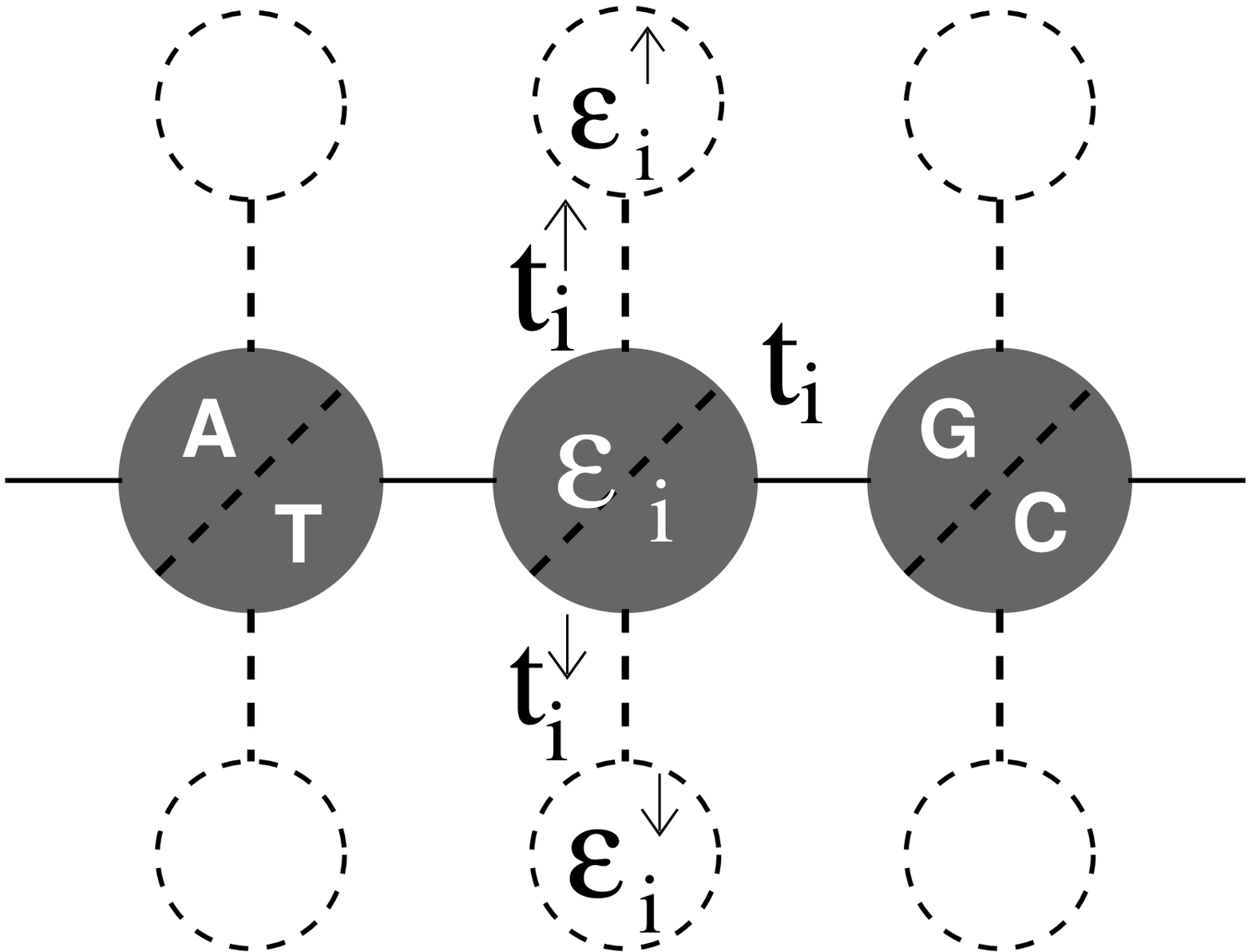}
  (b)\includegraphics[width=.21\textwidth]{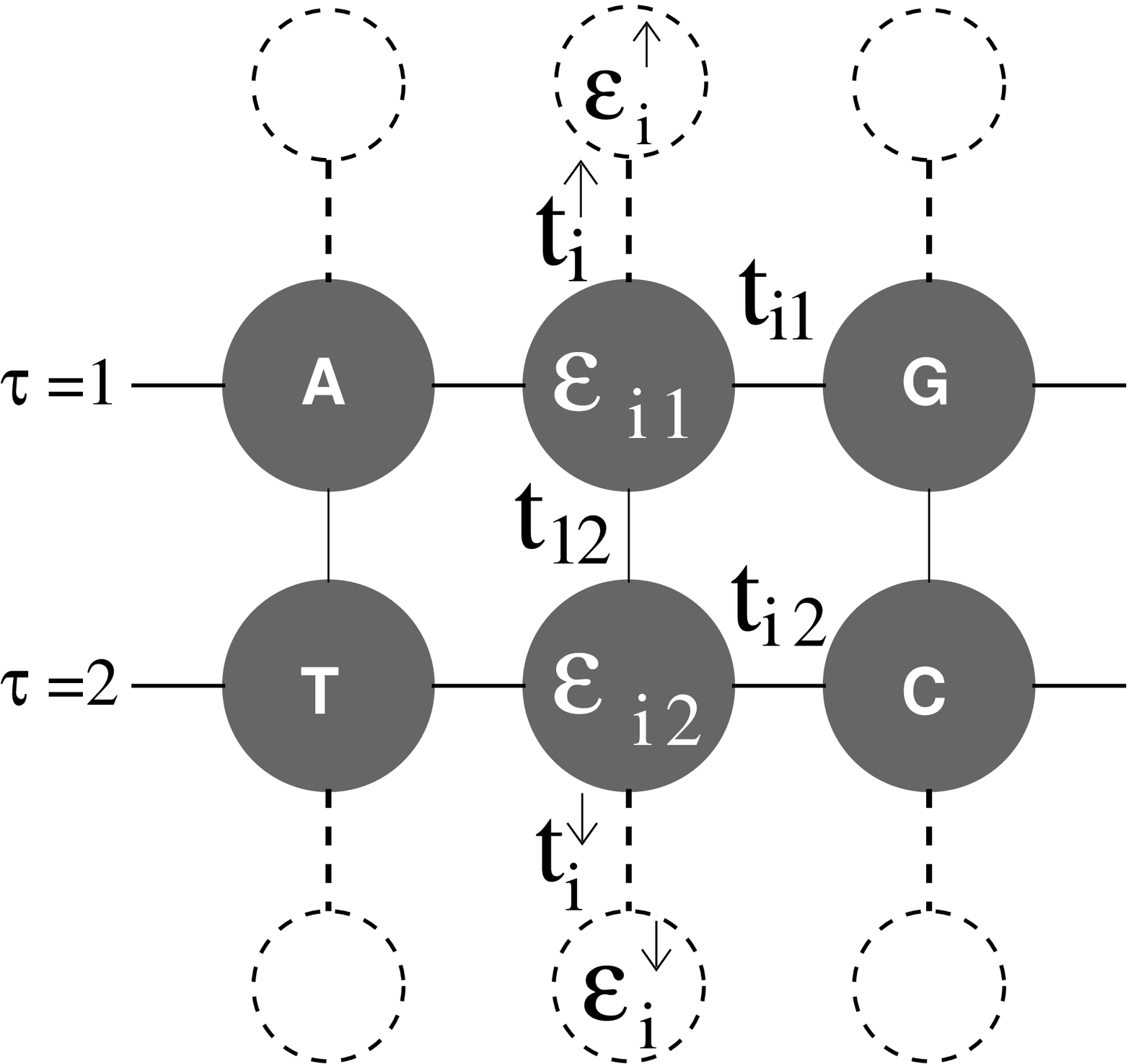}
  \caption{\label{fig:models} Schematic models for hole
    transport in DNA. The nucleobases are given as (grey) circles.
    Electronic pathways are shown as lines, and dashed lines and circles
    denote the sugar-phosphate backbone. Graph (a) shows effective
    models 1L and FB (with dashed backbone) for transport along a
    single channel, whereas graph (b) depicts possible two-channel
    transport models 2L and LM (with dashed backbone). }
\end{figure}

The generic form of the simple but physically
sounding 1-channel model of coherent hole transport of DNA is
given by an effective tight-binding Hamiltonian (the ``fishbone
model'' (FB)) \cite{CunCPD02}
\begin{eqnarray}
H_{\rm FB} &=& \sum_{i=1}^{L}
 \sum_{q=\uparrow,\downarrow} \left(
    -t_{i} |i \rangle \langle i+1|-t_i^q |i,q \rangle \langle i|
    \right.
 \nonumber \\
 & &
\left.+ \varepsilon_i |i \rangle \langle i| + \varepsilon_i^q |i,q
\rangle \langle i,q| \right) + h.c. \label{eq:fishbone-ham}
\end{eqnarray}
where each lattice point stands for a nucleotide base pair of the
chain for $i=1, \ldots, L$. $t_i$ is the hopping amplitude between
$i$th and $i+1$th base pairs and $\varepsilon_i$ is the on-site
potential of the $i$th base pair. $t_i^q$ with
$q=\uparrow,\downarrow$ is the hopping amplitude between the
$i$th base pair and its neighboring (upper and lower) backbone
sites $|i,q \rangle$. The onsite energy at the sites $|i,q
\rangle$ is given by $\varepsilon_i^q$. The model will be reduced
to the simplest one-ladder (1L) model if the sugar-phosphate
backbone sites $|i,q \rangle$ of DNA are absent, that is,
$t_i^q=\varepsilon_i^q=0$ \cite{BerBR01a,BerBR02,Roc03,RocBMK03}.
This one-channel model is shown schematically in Fig.\
\ref{fig:models}(a).

To account for the full double-strand nature of DNA, an
alternative two-channel ladder model (LM) shown in Fig.\
\ref{fig:models}(b) is also used. The corresponding Hamiltonian
is given as \cite{KloRT05}
\begin{eqnarray}
H_{\rm LM} &=& \sum_{i=1}^{L} \left[
 \sum_{\tau=1,2}
    \left( {t_{i,\tau}} |i,\tau\rangle \langle i+1,\tau| +
    \varepsilon_{i,\tau} |i,\tau\rangle \langle i,\tau| \right) \right.
 \nonumber \\
 & &  \mbox{} + \sum_{q=\uparrow,\downarrow}
    \left( \sum_{\tau=1.2} {t_i^q} |i,\tau\rangle \langle i,q(\tau)|+
    \varepsilon_i^q|i,q\rangle \langle i,q| \right)
\nonumber \\
 & & \mbox{ }
+ t_{1,2}|i,1\rangle \langle i,2| \Big]
 + h.c. \label{eq:ladder-ham}
\end{eqnarray}
where $t_{i,\tau}$ is the hopping amplitude between the sites
along each branch $\tau=1,2$ and $\varepsilon_{i,\tau}$ is the
corresponding onsite energy. $t_{12}$ represents the hopping
between the nucleotides of each base pair. Again, the model will
be reduced to a two-leg (2L) model if the backbone sites are not
taken into account \cite{Mal07,WanC06,GutMCP06}.


The onsite energies for each base are chosen according to the
ionization energies , $\epsilon_{\rm A}=8.24 e$V, $\epsilon_{\rm
C}=8.87 e$V, $\epsilon_{\rm G}=7.75 e$V and $\epsilon_{\rm T}=9.14
e$V \cite{SugS96,VoiJBR01,ZhaLHY02,YanWVW04,CauDL06} for each
model.
%
%
For model 1L, the hopping term between pairs base are all set as $t_n=0.4$ eV.
Other values ranging from $0.1$ to $1$ eV are also used to
investigate the robustness of our conclusion.
%
%
For model FB, $t_n$ is $0.4$ eV as in 1L. The additional hopping terms linking
to the backbone are taken as $0.7 e$V, whereas all backbone onsite
energies are assumed to be $8.5 e$V, roughly equal to the average
value of all onsite energies for the base pairs.
%
%
The hopping terms in model 2L between the same kind of base pairs (AT/AT,
GC/GC, etc.) are chosen as $0.35 e$V, and $0.17 e$V otherwise
\cite{CunCPD02}. Interchain coupling constant is fixed to
$t_{\perp}= 0.1 e$V.
%
%
Last, in model LM, intrachain and interchain hopping strengths are taken as in the
two-leg model 2L. Additionally, the backbone energetics is treated as
in the fishbone model case.

\section{Method}

The most convenient method to evaluate the transport properties of
these quasi-one-dimensional tight-binding models is known as the
transfer matrix method
(TMM) \cite{PicS81a,PicS81b,MacK83,KraM93,Mac94}. This approach
allows to determine the hole transmission coefficient $T(E)$ in
systems with varying cross section $M$ and length $L \gg M$. In
brief, the eigenstates $|\Psi\rangle = \sum_n \psi_n| n\rangle$
(here $|n\rangle$ denotes the $n$th site position of the hole) of the
Hamiltonian are computed from $\left(\psi_{L}, \psi_{L-1}\right)^T
= \tau_L\cdot \left( \psi_1, \psi_0 \right)^T$ where $\tau_L (E)$
is the global transfer matrix \cite{KraM93}. $E$ is the energy of
the injected carrier. The localization lengths are deduced from
the scaling analysis of $T(E)$, whatever the used effective model
\cite{PicS81a,PicS81b,MacK83,KraM93}. Besides, when assuming that
the DNA sequences are connected to the semi-infinite metallic
electrodes \cite{Cha07}, $T(E)$ takes the following analytical
form \cite{Roc03,ChaRS03,Mac99,Shi06,Shi06a}
\begin{equation}
T(E)=\frac{4-\bar{E}^2}{P+2-\bar{E}^2\tau_{11}\tau_{22}+\bar{E}(\tau_{11}-\tau_{22})(\tau_{12}-\tau_{21})}
\label{e:te}
\end{equation}
with $\bar{E}=(E-\varepsilon_m)/{t_0}$ and
$P=\sum_{i,j=1,2}\tau^2_{ij}$. $\varepsilon_m$ and $t_0$ are the
onsite energies and the hopping integral of the electrode
energetics, respectively. It is readily shown that
\begin{equation}
\tau_L=\left(\begin{array}{cc}\tau_{11} & \tau_{12}\\\tau_{21} &
\tau_{22}\end{array} \right)=M_LM_{L-1}\ldots M_2M_1
\end{equation}
with
\begin{equation}
M_n=\left(\begin{array}{cc}\frac{E-\epsilon_n}{t_n} &
-\frac{t_{n-1}}{t_{n}}\\1 & 0\end{array} \right)
\end{equation}
where $\epsilon_n=\varepsilon_n$ for 1L and $\varepsilon_n -
\sum_q \frac{(t_n^q)^2}{\varepsilon_n^q-E}$ for FB, respectively
\cite{KloRT05}. In the following
$\varepsilon_m=\varepsilon_G=7.75$ eV and $t_0=1$ eV.

To analyze the position-dependent transport properties of $p53$
gene and the effect of point mutations, let us define
$S=(s_1,s_2,\ldots,s_{20303})$ as a finite-length sequence of the
$p53$ gene. $S_{jL}$ is a segment of $S$ starting from the $j$th
base pair with length $L$, that satisfies $S_{jL}(n)=S(n+j-1)$
with $n=1,2,\cdots,L$. The transmission coefficient as a function
of energy is denoted as $T_{jL}(E)$. CT for the $j$th site with
propagation length $L$ is defined as the averaged value of the
integrated $T_{jL}(E)$ (for all incident energies) of all $L$
possible subsequences of $p53$ containing the $j$th site and with
length $L$
\begin{equation}
  \bar{T}_{j,L}=\frac{1}{L}\sum_{n=j-L+1}^{j}
  \frac{1}{E_1-E_0}\int^{E_1}_{E_0} T_{n,L}(E) dE
\label{eq:ave_te}.
\end{equation}
where $n$ is further restricted to $1\leq n \leq 20304-L$ close to
the boundaries; $E_0$ and $E_1$ denote a suitable
energy window which we shall normally choose to equal the extrema
of the energy spectrum for each model, i.e.\ $[6.5,10.5]$ for model 1L, $[7.5,10.5]$ for 2L, $[8,9.5]$ for FB and $[5,15]$ for LM.

If the $k$th base on the $p53$ sequence is mutated from $s_k$ to
$s$ and $j\le k\le j+L-1$ (i. e., the mutated site belongs to the
segment $S_{jL}$), the mutated sequence will be denoted as
$S^{ks}_{jL}$. $S^{ks}_{jL}(k-j+1)=s$ and $S^{ks}_{jL}(i\ne
k-j+1)=S_{jL}(i)$. The transmission coefficients of the original
and mutated sequences are denoted as $T_{jL}(E)$ and
$T^{ks}_{jL}(E)$, respectively.
The squared difference of the transmission coefficient between
the wild and mutated sequences is defined as
\begin{equation}
\Delta^{ks}_{jL}(E)\equiv \left[T_{jL}(E)-T^{ks}_{jL}(E)\right]^2 .
\label{e:delta}
\end{equation}
And $\Delta^{ks}_{jL}(E)$ is then summed for all incident energy $E$ as
\begin{equation}
\bar{\Delta}^{ks}_{jL}=\frac{1}{E_1-E_0}\int^{E_1}_{E_0}dE\Delta^{ks}_{jL}(E) .
\label{e:delta_ave}
\end{equation}
Finally, $\bar{\Delta}^{ks}_{jL}$ is averaged over all segments
with length $L$ containing the mutation site ($k$), to give the
average effect of the mutation $(k,s)$ on the change of CT for $p53$
\begin{equation}
\Gamma(k,s,L)=\frac{1}{L}\sum_{j}\bar{\Delta}^{ks}_{jL} .
\label{e:gamma}
\end{equation}
\begin{figure}
\centering
\includegraphics[width=.45\textwidth]{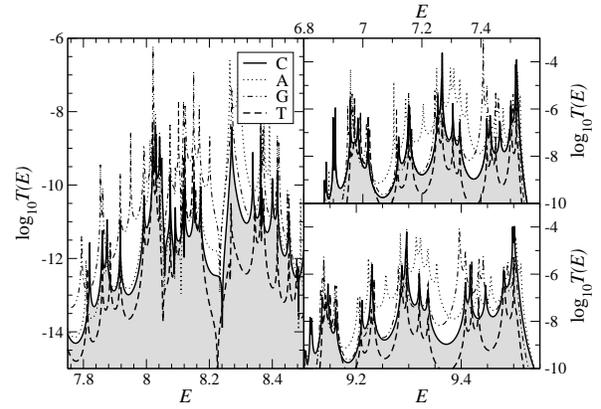}
\caption{ Energy-dependence of logarithmic transmission
coefficients $T^{14585,s}_{14570,31}(E)$ of the
original sequence ($C$ solid line) and mutated ($A$ dotted, $G$
dotted-dashed, $T$ dashed) sequences with length
$L=31$ (from $14570$th to $14500$th nucleotide) of $p53$. The left
panel shows results for model 2L, the right two panels denote the
two transport windows for the FB model\cite{KloRT05}.}
  \label{fig:mut}
\end{figure}
%
\begin{figure}
  \centering
  \includegraphics[width=.45\textwidth]{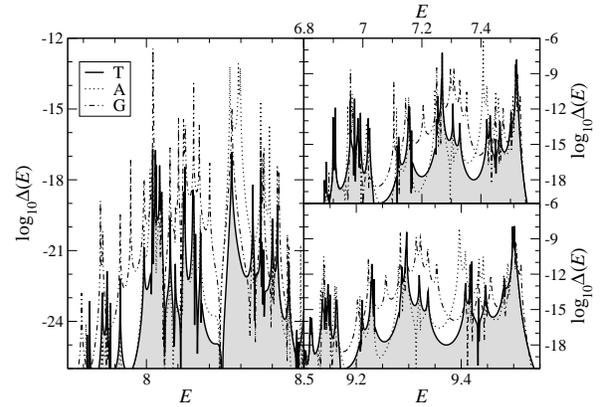}
  \caption{
    Energy-dependence of logarithmic squared differences
    $\Delta^{14585,s}_{14570,31}(E)$ between the transmission
    coefficients of the original sequence and mutated ($C\to T$ solid
line, $\to A$
    dotted, $\to G$ dotted-dashed) sequences. The
    left panel shows results for model 2L, the right panels
    denote model FB.}
  \label{fig:mut-diff}
\end{figure}

\section{Results~and~discussion}
The $14585$th base (exon 8, codon 306) of the $p53$ sequence is
found 133 times in the IARC database that mutates from $C$ to $T$
and causes various types of cancer \cite{why14585}. On the other
hand, the mutations $C \to G$ and $C\to A$ are said to be
noncancerous since they are never found in cancer cells. The
effects of the cancerous ($C\to T$) and noncancerous mutations
$T_{14570,31}^{14585,s}(E)$ and
$\Delta^{14585,s}_{14570,31}(E)$ for the models FB
and 2L are shown in \ref{fig:mut}, \ref{fig:mut-diff}
respectively. The overall effect of these three mutations
$\Gamma(14585,s,L=20, \ldots, 100)$ for all the $4$ models is
given in Table \ref{tab:gamma}.

It is clear from Table \ref{tab:gamma} that for
many cases the CT change due to cancerous mutation is much
smaller than noncancerous mutations. These results are stable over
a wide range of $L$ and model parameters. This suggests a scenario
to understand how specific mutation hotspots could be robust
against repair mechanism, and trigger carcinogenesis.
Experimentally, the BER enzymes can locate the damaged sites on
DNA by probing the CT of the segment bound by the enzymes
\cite{YavBSB05,YavSOD06}. If a mutation only weakly changes the
CT, the enzymes will not be able to find it and the repair
mechanism will not be activated. Such mutations will survive DNA
repair mechanims and yield cancers. In contrast, those mutations
that strongly affect CT could be more easily detected by the CT
probing mechanim of enzymes and therefore repaired.

The results presented thus far are for a particular hotspot. To
further challenge this scenario, many more hotspots of the $p53$ gene
have been analyzed. We have thus calculated $\bar{\Delta}(k,s,L)$ for $14$ hotspots with
the highest mutation frequencies and for $L$ up to $160$ in all $4$ models. The results show that the qualitative behavior of each hotspot for all models is similar. Thus the following analysis is performed on all hotspots for the 1L model.
\begin{table}
  \centering
\caption{
Renormalized values of the energy-averaged changes
$\bar{\Delta}^{14585,s}_{j,L}$ with
$L=11,21,\ldots,101$ and $j=14585-(L-1)/2$ in transmission
properties for the $4$ tight-binding models. All data are shown
with at most 3 significant figures. Common
    multiplication factors for each group of data for given $L$
    and mutations with $C\to A$, $G$ and $T$ are suppressed. Bold
    entries denote minima for the CT change of $C\to T$.}\label{tab:gamma}
  \begin{tabular*}{\hsize}{@{\extracolsep{\fill}}cccccc}
    \hline\hline
    $s$      & $L$& 1L          & FB          & 2L         & LM \\ \hline
    $C\to A$ & 11 & 71.2        & 1.276       & 4.84       & 3.24 \\ \hline
    $C\to G$ & 11 & 113         & 0.164       & 5.70       & 5.40 \\ \hline
    $C\to T$ & 11 & {\bf 16.3}  & {\bf 0.013} & {\bf 1.62} & {\bf 0.29}
\\ \hline \hline
    $C\to A$ & 21 & 16.4        & 7.58        & 2.70       & 5.31 \\ \hline
    $C\to G$ & 21 & 30.5        & 1.08        & 5.52       & 44.3 \\ \hline
    $C\to T$ & 21 & {\bf 3.23}  & {\bf 0.19}  & {\bf 0.18} & {\bf 3.73} \\ \hline
\hline
    $C\to A$ & 31 & 15.7        & 548         & 14.6       & 13.9 \\ \hline
    $C\to G$ & 31 & 21.4        & 5459        & 5.18       & 0.55 \\ \hline
    $C\to T$ & 31 & {\bf 9.14}  & {\bf 0.63  }& {\bf 3.60} & {\bf 0.23}
\\ \hline \hline
    $C\to A$ & 41 & 1.16        & 30.7        & 0.52       & 3.66 \\ \hline
    $C\to G$ & 41 & 2.21        & 0.72        & 2.99       & 5.23 \\ \hline
    $C\to T$ & 41 & {\bf 0.40}  & {\bf 0.009} &      1.36  & {\bf 1.17}
\\ \hline \hline
    $C\to A$ & 51 & 0.56        & 232         & 1.60       & 1.00 \\ \hline
    $C\to G$ & 51 & 0.90        & 0.21        & 41.7       & 2527 \\ \hline
    $C\to T$ & 51 & {\bf 0.71}  & {\bf 0.13 } &      3.36  & 9.56
\\ \hline \hline
    $C\to A$ & 61 & 0.84        & 3160        & 1.50       & 0.70 \\ \hline
    $C\to G$ & 61 & 2581        & 2.95        & 1.26       & 9.01 \\ \hline
    $C\to T$ & 61 & 1.29        & {\bf 1.84 } &      14.4  & 99.0
\\ \hline \hline
    $C\to A$ & 71 & 0.99        & 3187        & 1.48       & 0.12 \\ \hline
    $C\to G$ & 71 & 9.03        & 0.29        & 1.45       & 0.91 \\ \hline
    $C\to T$ & 71 & 4.59        & {\bf 0.19 } &      1.47  &      18.4
\\ \hline \hline
    $C\to A$ & 81 & 3.61        & 3939        & 5.53       & 3.19 \\ \hline
    $C\to G$ & 81 & 237         & 3.61        & 5.49       & 5.48 \\ \hline
    $C\to T$ & 81 & {\bf 0.14}  & {\bf 2.30}  & {\bf 5.40} & {\bf 0.90}
\\ \hline \hline
    $C\to A$ & 91 & 1.06        & 1183        & 10.1       & 11.5 \\ \hline
    $C\to G$ & 91 & 232         & 1.1         & 92.6       & 60.4 \\ \hline
    $C\to T$ & 91 & 2.95        & {\bf 0.69 } & {\bf 0.32} & {\bf 0.47}
\\ \hline \hline
    $C\to A$ &101 & 1.63        & 9143        & 199.1      & 102.2\\ \hline
    $C\to G$ &101 & 1044        & 8.68        & 820.1      & 493.3\\ \hline
    $C\to T$ &101 & 8.64        & {\bf 5.33}  & {\bf 0.1 } & {\bf 0.1 }
\\ \hline \hline
\end{tabular*}
\end{table}
%
\begin{figure}
\centering
\includegraphics[angle=-90,width=.45\textwidth]{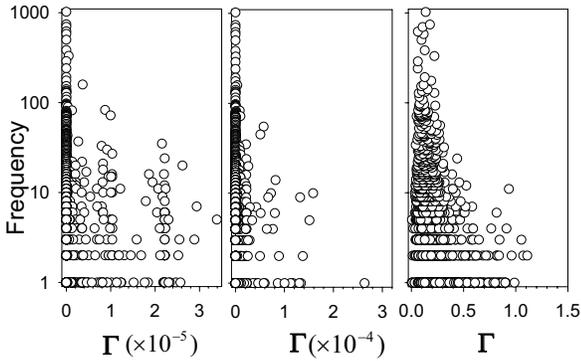}
\caption{$\Gamma(k,s,L)$ for 1L model with (a) $(t_n,L)=(0.1,20)$,
(b) $(0.4,80)$, and (c) $(1.0, 140)$ for all cancerous point
mutations of $p53$ and their frequencies found in the IARC
database.} \label{fig:scatt}
\end{figure}
Fig.\ \ref{fig:scatt}(b) shows the correlation between frequency
found in the cancer cells and the CT change $\Gamma(k,s,L=80)$
with $t_n=0.4$ eV. It is clear that the hotspots with highest
frequencies correspond to smaller $\Gamma$. Thus the correlation
observed in Fig.\ \ref{fig:mut} for the $14585$th
site is common for most of the hotspots. Fig.\ \ref{fig:scatt}(a)
and (c) show similar behaviors for $t_n=0.1$ and $1$ eV,
respectively. The scenario is thus found to be robust for a wide
range of $t_n$.

\section{Conclusion}

The CT modifications due to all possible point
mutations of the $p53$ tumor suppressor gene have been analyzed by
TMM together with statistical methods. The results show that on
average the cancerous mutations of the gene yield smaller changes
of the CT in contrast with non-cancerous mutations. The tendency
is valid for the $4$ studied tight-binding models (1L, FB, 2L,
and LM) and is robust for a wide range of the hopping integral
$t_n$ ($0.1\sim 1.0$ eV).

These results suggest a possible scenario of how cancerous
mutations might circumvent the DNA damage-repair mechanism and
survive to yield carcinogenesis. However, our analysis is only
valid in a statistical sense and we do observe occasional
non-cancerous mutations with weak chan\-ge of CT. For these, other
DNA repair processes should exist and we therefore do not intend
to claim that the DNA-damage repair solely uses a CT-based
criterion. Still, our results exhibit an intriguing and new
correlation between the electronic structure of DNA hotspots and
the DNA da\-ma\-ge-repair process.

One notes that to further support the abovementionned scenario,
additional complexities of the DNA energetics should also be
considered. This includes  to investigate the role of
electron-phonon coupling, polaronic transport, more detailed
sequence-dependent energetics such as two-strand couplings,
electronic correlations, as well as metal/DNA contact interactions
\cite{DiaSSD07,KloRT05,Mal07,WanC06,GutMCP06,Sen03,ConR00,Con05,WeiWCY05,Sta03,Mac05}.
Ultimately, experimental studies of short strands of wild and
mutated subsequences of the p53 gene should be performed to
challenge our theory. The lengths scales of DNA required to unveil
our mechanism are already within the scope of experimental
measurements \cite{PorBVD00,XuZLT04,CohNNP05,KasKGR01}.

\begin{acknowledgement}
We thank J.\ Cadet, D.\ Gasparutto, Th.\ Douki, M. F. Yang, and
M.\ Turner for useful discussions. This work is partially
supported by the National Science Council and National Center for
Theoretical Sciences in Taiwan (CTS, grant no.\
95-2112-M-029-003-) and the UK Leverhulme Trust (RAR, grant no.\
F/00 215/AH).  Part of the calculations were performed in the
National Center for High-Performance Computing in Taiwan.
\end{acknowledgement}

%
%










\begin{thebibliography}{[10]}

\bibitem{EndCS04}
 \textsc{R.\,G. Endres},  \textsc{D.\,L. Cox},  and  \textsc{R.\,P.
  Singh},
 \jr{Rev. Mod. Phys.} \textbf{76}, 195--214 (2004).


\othercit
\bibitem{Cha07}
 \textsc{T.~Chakraborty} (ed.),
Charge Migration in DNA: Perspectives from Physics, Chemistry and
Biology
  (Springer Verlag, Berlin, 2007).


\othercit
\bibitem{GutC07}
 \textsc{R.~Gutierrez} and  \textsc{G.~Cuniberti},
NanoBioTechnology: BioInspired device and materials of the future,
 (Humana Press, Totowa, 2007), chap. Hamiltonian approaches to charge transport
  in {DNA} molecular wires.


\othercit
\bibitem{BerKBR04}
 \textsc{Y.\,A. Berlin},  \textsc{I.\,V. Kurnikov},  \textsc{D.\,N. Beratan},
  \textsc{M.\,A. Ratner},  and  \textsc{A.\,L. Burin},
Topics in Current Chemistry,
 (Springer, Berlin, 2004),  pp.\,1--36.


\bibitem{RajJB00}
 \textsc{S.\,R. Rajski},  \textsc{B.\,A. Jackson},  and  \textsc{J.\,K.
  Barton},
 \jr{Mutat. Res.} \textbf{447}, 49--72 (2000).


\bibitem{YavBSB05}
 \textsc{E.~Yavin},  \textsc{A.\,K. Boal},  \textsc{E.\,D.\,A. Stemp},
  \textsc{E.\,M. Boon},  \textsc{A.\,L. Livingston},  \textsc{V.\,L.\,O. Shea},
   \textsc{S.\,S. David},  and  \textsc{J.\,K. Barton},
 \jr{Proc. Natl. Acad. Sci.} \textbf{102}, 3546 (2005).


\bibitem{YavSOD06}
 \textsc{E.~Yavin},  \textsc{E.\,D.\,A. Stemp},  \textsc{V.\,L. O'Shea},
  \textsc{S.\,S. David},  and  \textsc{J.\,K. Barton},
 \jr{Proc. Nat. Acad. Sci.} \textbf{103}, 3610 (2007).


\bibitem{LofP96}
 \textsc{S.~Loft} and  \textsc{H.\,E. Poulsen},
 \jr{J. Mol. Med.} \textbf{74}, 297--312 (1996).


\bibitem{Lan92}
 \textsc{D.\,P. Lane},
 \jr{Nature} \textbf{358}, 15 (1992).


\bibitem{She04}
 \textsc{C.\,J. Sherr},
 \jr{Cell} \textbf{234}, 116 (2004).


\bibitem{PetMKI07}
 \textsc{A.~Petitjean},  \textsc{E.~Mathe},  \textsc{S.~Kato},
  \textsc{C.~Ishioka},  \textsc{S.~Tavtigian},  \textsc{P.~Hainaut},  and
  \textsc{M.~Olivier},
 \jr{Hum. Mutat.} \textbf{28}(6), 622--29 (2007),
http://www-p53.iarc.fr/index.html, R11.


\bibitem{CunCPD02}
 \textsc{G.~Cuniberti},  \textsc{L.~Craco},  \textsc{D.~Porath},  and
  \textsc{C.~Dekker},
 \jr{Phys. Rev. B} \textbf{65}, 241314--4 (2002).


\bibitem{Sta05}
 \textsc{E.\,B. Starikov},
 \jr{Phil. Mag.} \textbf{85}(29), 3435--3462 (2005).


\bibitem{BerBR01a}
 \textsc{Y.\,A. Berlin},  \textsc{A.\,L. Burin},  and  \textsc{M.\,A.
  Ratner},
 \jr{Superlattices and Microstructures} \textbf{28}(4), 241--252 (2001).


\bibitem{BerBR02}
 \textsc{Y.\,A. Berlin},  \textsc{A.\,L. Burin},  and  \textsc{M.\,A.
  Ratner},
 \jr{Chem. Phys.} \textbf{275}, 61--74 (2002).


\bibitem{Roc03}
 \textsc{S.~Roche},
 \jr{Phys. Rev. Lett.} \textbf{91}, 108101--4 (2003).


\bibitem{RocBMK03}
 \textsc{S.~Roche},  \textsc{D.~Bicout},  \textsc{E.~{Maci\'{a}}},  and
  \textsc{E.~Kats},
 \jr{Phys. Rev. Lett.} \textbf{91}, 228101--4 (2003).


\bibitem{KloRT05}
 \textsc{D.\,K. Klotsa},  \textsc{R.\,A. {R\"{o}mer}},  and  \textsc{M.\,S.
  Turner},
 \jr{Biophys. J.} \textbf{89}, 2187--2198 (2005).


\bibitem{Mal07}
 \textsc{A.\,V. Malyshev},
 \jr{Phys. Rev. Lett.} \textbf{98}(9), 096801 (2007).


\bibitem{WanC06}
 \textsc{X.\,F. Wang} and  \textsc{T.~Chakraborty},
 \jr{Phys. Rev. Lett.} \textbf{97}, 106602 (2006).


\bibitem{GutMCP06}
 \textsc{R.~Gutierrez},  \textsc{S.~Mohapatra},  \textsc{H.~Cohen},
  \textsc{D.~Porath},  and  \textsc{G.~Cuniberti},
 \jr{Phys. Rev. B} \textbf{74}, 235105 (2006).


\bibitem{SugS96}
 \textsc{H.~Sugiyama} and  \textsc{I.~Saito},
 \jr{J. Am. Chem. Soc.} \textbf{118}(30), 7063--7068 (1996).


\bibitem{VoiJBR01}
 \textsc{A.\,A. Voityuk},  \textsc{J.~Jortner},  \textsc{M.~Boxin},  and
  \textsc{N.~R\"{o}sch},
 \jr{J. Chem. Phys.} \textbf{114}(13), 5614--5620 (2001).


\bibitem{ZhaLHY02}
 \textsc{H.~Zhang},  \textsc{X.\,Q. Li},  \textsc{P.~Han},  \textsc{X.\,Y. Yu},
   and  \textsc{Y.~Yan},
 \jr{Journal of Chemical Physics} \textbf{117}(9), 4578--4584 (2002).


\bibitem{YanWVW04}
 \textsc{X.~Yang},  \textsc{X.\,B. Wang},  \textsc{E.\,R. Vorpagel},  and
  \textsc{L.\,S. Wang},
 \jr{Proc. Nat. Acad. Sci.} \textbf{101}(51), 17588--17592 (2004).


\bibitem{CauDL06}
 \textsc{E.~Cauet},  \textsc{D.~Dehareng},  and  \textsc{J.~Lievin},
 \jr{J.\ Phys.\ Chem.\ A} \textbf{110}(29), 9200--9211 (2006).


\bibitem{PicS81a}
 \textsc{J.\,L. Pichard} and  \textsc{G.~Sarma},
 \jr{J. Phys. C} \textbf{14}, L127--L132 (1981).


\bibitem{PicS81b}
 \textsc{J.\,L. Pichard} and  \textsc{G.~Sarma},
 \jr{J. Phys. C} \textbf{14}, L617--L625 (1981).


\bibitem{MacK83}
 \textsc{A.~MacKinnon} and  \textsc{B.~Kramer},
 \jr{Z. Phys. B} \textbf{53}, 1--13 (1983).


\bibitem{KraM93}
 \textsc{B.~Kramer} and  \textsc{A.~MacKinnon},
 \jr{Rep. Prog. Phys.} \textbf{56}, 1469--1564 (1993).


\bibitem{Mac94}
 \textsc{A.~MacKinnon},
 \jr{J. Phys.: Condens. Matter} \textbf{6}, 2511--2518 (1994).


\bibitem{ChaRS03}
 \textsc{A.~Chakrabarti},  \textsc{R.\,A. R\"{o}mer},  and
  \textsc{M.~Schreiber},
 \jr{Phys. Rev. B} \textbf{68}(19), 195417 (2003).


\bibitem{Mac99}
 \textsc{E.~Maci\'a},
 \jr{Phys. Rev. B} \textbf{60}(14), 10032--10036 (1999).


\bibitem{Shi06}
 \textsc{C.\,T. Shih},
 \jr{phys. stat. sol. (b)} \textbf{243}(2), 378--381 (2006).


\bibitem{Shi06a}
 \textsc{C.\,T. Shih},
 \jr{Phys. Rev. E} \textbf{74}(1), 010903 (2006).

\bibitem{why14585}All the three types of mutations on the positions of
the eight most frequent mutations are cancerous. So we choose this
$9-th$ hotspot because it has two non-cancerous mutations for
comparison.


\bibitem{DiaSSD07}
 \textsc{E.~D\'{i}az},  \textsc{A.~Sedrakyan},  \textsc{D.~Sedrakyan},  and
  \textsc{F.~Dom\'{i}nguez-Adame},
 \jr{Phys. Rev. B} \textbf{75}, 014201 (2007).


\bibitem{Sen03}
 \textsc{K.~Senthilkumar},  \textsc{F.\,C. Grozema},  \textsc{C.\,.\,F.
  Guerra},  \textsc{F.\,M. Bickelhaupt},  and  \textsc{L.\,D.\,A. Siebbeles},
 \jr{J. Am. Chem. Soc.} \textbf{125}, 13658--13659 (2003).


\bibitem{ConR00}
 \textsc{E.\,M. Conwell} and  \textsc{S.\,V. Rakhmanova},
 \jr{Proc. Nat. Acad. Sci.} \textbf{97}(9), 4556--4560 (2000).


\bibitem{Con05}
 \textsc{E.\,M. Conwell},
 \jr{Proc. Nat. Acad. Sci.} \textbf{102}, 8795--8799 (2005).


\bibitem{WeiWCY05}
 \textsc{J.\,H. Wei},  \textsc{L.\,X. Wang},  \textsc{K.\,S. Chan},  and
  \textsc{Y.~Yan},
 \jr{Phys. Rev. B} \textbf{72}(6), 064304 (2005).


\bibitem{Sta03}
 \textsc{E.\,B. Starikov},
 \jr{Phil. Mag. Lett.} \textbf{83}, 699 (2003).


\bibitem{Mac05}
 \textsc{E.~Macia},  \textsc{F.~Triozon},  and  \textsc{S.~Roche},
 \jr{Phys. Rev. B} \textbf{71}, 113106 (2005).


\bibitem{PorBVD00}
 \textsc{D.~Porath},  \textsc{A.~Bezryadin},  \textsc{S.~Vries},  and
  \textsc{C.~Dekker},
 \jr{Nature} \textbf{403}, 635--638 (2000).


\bibitem{XuZLT04}
 \textsc{B.~Xu},  \textsc{P.~Zhang},  \textsc{X.~Li},  and
  \textsc{N.~Tao},
 \jr{Nano Lett.} \textbf{4}, 1105 (2004).


\bibitem{CohNNP05}
 \textsc{H.~Cohen},  \textsc{C.~Nogues},  \textsc{R.~Naaman},  and
  \textsc{D.~Porath},
 \jr{Proc. Nat. Acad. Sci.} \textbf{102}, 11589 (2005).


\bibitem{KasKGR01}
 \textsc{A.\,Y. Kasumov},  \textsc{M.~Kociak},  \textsc{S.~Gu{\'e}ron},
  \textsc{B.~Reulet},  \textsc{V.\,T. Volkov},  \textsc{D.\,V. Klinov},  and
  \textsc{H.~Bouchiat},
 \jr{Science} \textbf{291}, 280--282 (2001).


\end{thebibliography}

\providecommand{\WileyBibTextsc}{}
\let\textsc\WileyBibTextsc
\providecommand{\othercit}{} \providecommand{\jr}[1]{#1}
\providecommand{\etal}{~et~al.}

\end{document}